\documentstyle[12pt,epsfig,amssymb]{article}
\begin{document}
\setlength{\parskip}{0.45cm}
\setlength{\baselineskip}{0.75cm}
\begin{titlepage}
\begin{flushright}
CERN-TH/98-201\\
CPT-98/PE.3666\\
DTP/98/38\\
hep-ph/9806513\\ 
June 1998
\end{flushright}
\vspace{0.3cm}
\begin{center}
\Large
{\bf Bounds on Transverse Spin Asymmetries for}

\vspace{0.2cm}
{\bf $\Lambda$ Baryon Production in $pp$ Collisions at BNL RHIC}

\vspace{0.8cm}
\large
D.\ de Florian$^{a}$, J.\ Soffer$^{b}$,  M.\ Stratmann$^{c}$, and 
W.\ Vogelsang$^{a}$

\vspace{0.6cm}
\normalsize
$^a$Theoretical Physics Division, CERN, CH-1211 Geneva 23, Switzerland

\vspace{0.1cm}
$^b$Centre de Physique Th\'eorique CNRS Luminy Case 907,\\            
F-13288 Marseille Cedex 09, France

\vspace{0.1cm}
$^c$Department of Physics, University of Durham, Durham DH1 3LE, England

\vspace{0.6cm}
\large
{\bf Abstract} 
\end{center}
\vspace{0.1cm}
\noindent
We study inclusive $\Lambda$ hyperon production in $pp$ collisions at 
BNL RHIC, with just one transversely polarized proton. We show that the 
measurement of the spin transfer between the initial proton and the 
produced $\Lambda$ is sensitive to the proton transversity distributions 
and to the corresponding $\Lambda$ transversity fragmentation functions. 
In view of our present ignorance of these distributions and 
fragmentation functions, we resort to positivity bounds for making some 
predictions for the corresponding spin transfer asymmetry.
\vspace{0.5cm}
\begin{flushleft}
\end{flushleft}
\end{titlepage}
%
%
\noindent
{}From longitudinally polarized deep inelastic scattering (DIS) experiments, 
we only begin to gain some insight into the helicity parton densities 
$\Delta_L f(x,Q^2)$, with $f=q,\,\bar{q},\,g$. Due to the scarcity of the 
data and their limited kinematical coverage in $x$ and $Q^2$,
many uncertainties remain though, in particular concerning the 
precise $x$--shape of the polarized gluon distribution and the
flavour decomposition of the quark singlet combination.
Hence the r\^ole played by (anti)quarks and gluons in the nucleon 
spin sum rule is still unsettled.
To reduce our present ignorance, it is desirable to 
study polarization effects also in various other processes, rather than in 
fully inclusive DIS. This will be achieved by the vast spin physics 
programme which will be undertaken at the forthcoming RHIC collider 
at BNL \cite{ref:rhic}.
Of particular interest here are also reactions with a detected 
(longitudinally) polarized hadron $h$ in the
final state which would allow us to study spin-dependent 
{\em fragmentation} functions $\Delta_L D_f^h(x,Q^2)$, the even far 
less well-determined ``time-like'' counterparts of the parton density 
functions $\Delta_L f(x,Q^2)$.
In addition, such measurements could provide a deeper 
understanding of the {\em hadronization} mechanism for  
hadrons produced inclusively.
So far, the only measured polarized
hadron in the final state is the $\Lambda$ hyperon. 
Its $\Delta_L D_f^{\Lambda}(x,Q^2)$ are only poorly constrained 
by the existing $e^+e^- \rightarrow \vec{\Lambda} X$ data \cite{ref:dsvlambda}, 
taken on the $Z$-resonance (henceforth, a horizontal (vertical)
arrow will denote a longitudinally (transversely) polarized particle). 
However, it was shown in a recent work \cite{ref:dsvrhic}, that  
studying the helicity transfer between the initial proton and a high-$p_T$
$\Lambda$ in $\vec{p} p\to \vec{\Lambda} X$
 would allow us to 
discriminate cleanly between various 
possible scenarios for $\Delta_L D_f^{\Lambda}(x,Q^2)$ proposed in
\cite{ref:dsvlambda}.

In the case of {\em transverse} polarization the situation is even worse. The
transversity densities, denoted by $\Delta_T q(x,Q^2)$
(or $h_1^q(x,Q^2)$), which are equally fundamental at leading twist 
\cite{ref:jjrp} as the $\Delta_L q(x,Q^2)$, are experimentally
completely unknown for the time being. 
The chiral-odd $\Delta_T q(x,Q^2)$ measure the difference of the probabilities 
to find a quark with its spin parallel to that of a transversely polarized
nucleon and of finding it oppositely polarized. 
Unlike the case of unpolarized and longitudinally polarized densities,
there is no gluon transversity distribution at leading twist due to angular
momentum conservation \cite{ref:jjrp,ref:jmj}, and the $\Delta_T q(x,Q^2)$
are not accessible in inclusive DIS measurements because of their
chirality properties.
Various ways to measure the $\Delta_T q(x,Q^2)$ have been suggested,
for instance via the transversely polarized Drell-Yan process 
\cite{ref:jjrp,ref:dyold,ref:mssv} at RHIC, but, as already mentioned, 
no data are available so far. 
In a similar way, one can define transversity fragmentation functions, 
denoted by $\Delta_T D_q^h (x,Q^2)$, to describe the fragmentation of a 
transversely polarized quark into a transversely polarized hadron. 
Needless to say that also in this case, we have no experimental information 
on these  quantities. 
In view of the promising results obtained in \cite{ref:dsvrhic} 
concerning a possible measurement of the $\Delta_L D_f^{\Lambda}(x,Q^2)$ 
in $\vec{p} p\to \vec{\Lambda} X$, 
it seems worthwhile to study this reaction for the situation of 
transverse polarization at RHIC, i.e., for 
$p^{\uparrow} p\to \Lambda^{\uparrow} X$. 
This is the main purpose of this paper. 
In order to be able to make sensible 
predictions for the possible spin-transfer asymmetries for this process, 
we will exploit the positivity constraints derived in \cite{ref:js} to
constrain the involved quantities $\Delta_T q(x,Q^2)$ and 
$\Delta_T D_q^h (x,Q^2)$ in a non-trivial way.

%
Let us first recall that a positivity constraint at the naive 
parton model level was obtained for the $\Delta_T q(x)$, 
which reads \cite{ref:js}
\begin{equation}
\label{eq:pos}
2|\Delta_T q(x)| \le q(x) + \Delta_L q(x) \; .
\end{equation}
This result follows from the positivity properties of the forward 
quark-nucleon elastic amplitude, for which $\Delta_T q$ corresponds to
\begin{equation}
\label{eq:ampl}
q_{h'} (q) + N_H (P) \to q_{h} (q) + N_{H'} (P) \; ,
\end{equation}
where the helicities are such that $H=h=+1/2$ and $H'=h'=-1/2$. 
It was shown recently that Eq.~(\ref{eq:pos}) is preserved by the 
QCD $Q^2$ evolution, even to next-to-leading order (NLO) accuracy
\cite{ref:mssv,ref:wv,ref:bst}. Eq.~(\ref{eq:ampl}) as it stands only 
applies to the emission of a quark by a nucleon, but by using time 
reversal it is also related to the fragmentation of a quark into a nucleon.
Here, keeping the same helicity labels as above, it corresponds to 
$\Delta_T D_q^h(x)$. Consequently an analogous positivity bound for the 
fragmentation functions of a quark $q$ into a hadron $h$ holds, namely
\begin{equation}
\label{eq:posfrag}
2|\Delta_T D_q^h (x)| \le  D_q^h (x) + \Delta_L D_q^h (x) \; .
\end{equation}
This new result is surely valid at the level of the naive parton model, and
below, after specifying the densities on the r.h.s.\ of
Eq.~(\ref{eq:posfrag}), we will show that it is also 
maintained by the QCD $Q^2$ evolution at leading order (LO).
We will use these non-trivial bounds (\ref{eq:pos}) and (\ref{eq:posfrag}) to 
constrain the unmeasured transversity parton densities $\Delta_T q(x,Q^2)$
and fragmentation functions $\Delta_T D_q^h (x,Q^2)$ in our studies
of the spin-transfer asymmetry for transversely polarized $\Lambda$ 
baryon production at RHIC below.  

%
But before going into the details, we recall some general positivity 
constraints 
at the hadronic level that, even though not serving as a further constraint
on our parton densities and fragmentation functions, will provide a 
consistency check on our calculation.
A reaction of the type $pp \to \Lambda X$, where only one 
initial proton and the $\Lambda$ are polarized, 
can be described in terms of seven spin observables \cite{ref:dm}.
By studying the positivity domain of the reaction, 
one finds several model-independent constraints 
among these spin observables, which are valid at any kinematical point 
(total energy, transverse momentum, rapidity, etc.). If we restrict ourselves 
to the observables calculable in QCD at leading twist, only three of these spin 
transfer parameters survive, which can be chosen to be the spin transfer
asymmetries $D_{LL}$, $D_{SS}$ and $D_{NN}$. Here ``$L$'' stands for 
longitudinal polarization of the proton and the $\Lambda$, whereas ``$S$'' and 
``$N$'' denote transverse polarization, with the proton and the $\Lambda$ 
polarization vectors in, or normal to, the scattering plane, respectively.  
For all three cases one has the usual definition of a spin transfer asymmetry, 
\begin{equation} 
\label{eq:asydef}
D_{PP} \equiv \frac{\sigma (s_p,s_{\Lambda}) - \sigma (s_p,-s_{\Lambda})}
{\sigma (s_p,s_{\Lambda}) + \sigma (s_p,-s_{\Lambda})} \; , \; \; \; \; 
(P=L,S,N) \; ,
\end{equation}
where $s_p$, $s_{\Lambda}$ are the proton and $\Lambda$ spin vectors. In each
of the cases $P=L,S,N$, the sum in the denominator of (\ref{eq:asydef}) 
corresponds to the usual unpolarized cross section for $\Lambda$ production
in $pp$ scattering. As can be derived from \cite{ref:dm}, 
the $D_{PP}$ are subject to the following constraint:
\begin{equation}
\label{eq:doncel}
|D_{LL} \pm D_{SS}| \leq 1 \pm D_{NN} \; . 
\end{equation}
For the process $p^{\uparrow} p\to \Lambda^{\uparrow} X$ considered here
at LO QCD, it will actually turn out that $D_{SS} \equiv D_{NN}$ (see below). 
Eq.~(\ref{eq:doncel}) therefore reduces to
\begin{equation}
\label{eq:doncel2}
|D_{LL} \pm D_{NN}| \leq 1 \pm D_{NN} \; . 
\end{equation}
Since we are left with essentially only one independent transverse-spin 
observable, we will refer to it by the label ``$T$'' from now on and abbreviate
the r.h.s.\ of Eq.~(\ref{eq:asydef}) as
\begin{equation} 
\label{eq:asydef1}
D_{PP} \equiv \frac{\Delta_P \sigma}{\sigma} \; ,
\end{equation}
where $P=L,T$. The case $P=L$ was already studied in detail in \cite{ref:dsvrhic}; 
as mentioned above, the present paper deals with $P=T$. Here we will closely 
follow the procedure adopted in \cite{ref:dsvrhic}.

In various theoretical analyses of spin-transfer reactions it has turned 
out to be particularly useful to study distributions differential in the 
rapidity of a produced particle, see, e.g., \cite{ref:dsvrhic}, 
to which we therefore limit 
ourselves also in the present analysis. The rapidity differential polarized 
cross section in the numerator of (\ref{eq:asydef1}) can be schematically 
written in a factorized form as
\begin{eqnarray}
\label{eq:cross}
\nonumber
\frac{d\Delta_P \sigma^{pp\rightarrow \Lambda X}}{d \eta} = \\
&&\hspace*{-3.1cm} \int_{p_T^{min}} \hspace{-0.2cm} \!\! dp_T 
\sum_{ff'\rightarrow i X'} 
\int dx_1 dx_2 dz\, f^p(x_1,\mu^2) \times \Delta_P f'^p(x_2,\mu^2) \times
\Delta_P D_i^{\Lambda}(z,\mu^2) \times 
\frac{d\Delta_P \hat{\sigma}}{d\eta} \; ,
\end{eqnarray}
the sum running over all possible LO subprocesses $ff'\rightarrow iX'$ 
(partons $f'$ and $i$ are polarized) with spin-transfer 
cross sections $d\Delta_P \hat{\sigma}/d\eta$ defined in complete
analogy with the numerator of Eq.~(\ref{eq:asydef}). Note the appearance of the 
usual unpolarized parton densities $f^p$ in (\ref{eq:cross}), resulting from the 
fact that one initial proton is unpolarized. The expression for the 
unpolarized cross section $d\sigma^{pp\rightarrow \Lambda X}/d \eta$, needed
to calculate the spin-transfer asymmetries in (\ref{eq:asydef}), is similar
to the one in (\ref{eq:cross}), with all $\Delta$'s removed.
In (\ref{eq:cross}), we have integrated over the transverse momentum $p_T$ 
of the $\Lambda$, with $p_T^{min}$ denoting some suitable lower cut-off to 
be specified below.

The spin-transfer cross sections for the subprocesses $ff'\rightarrow 
iX'$ have been known for quite some time. 
They can be found in \cite{ref:handedness}
for both polarization cases ($P=L,T$). The cross sections for the 
transversity case, $P=T$, were presented in \cite{ref:handedness} 
in a form that 
also allows us to distinguish between the situations ``$S$'' and ``$N$'' introduced
above, i.e., when the final-state particle ``$i$'', and hence the $\Lambda$, is
transversely polarized in, or normal to, the scattering plane: writing the 
momentum of particle ``$i$'' in terms of its transverse momentum $p_T^i
\equiv p_T/z$, its pseudorapidity $\eta$ and its azimuthal angle $\Phi$ as   
\begin{equation} 
\vec{p}_{i}=p_T^i (\cos\Phi,\sin\Phi,\sinh \eta) \; , 
\end{equation}
one can parametrize~\cite{ref:handedness} the transverse spin vector of the 
$\Lambda$ by 
\begin{equation}
\label{eq:sc}
\vec{s}_{\Lambda}(\beta)=(\sin\Phi \cos\beta+\tanh \eta \cos\Phi\sin\beta,
-\cos\Phi\cos\beta+\tanh \eta \sin\Phi\sin\beta,-\sin\beta/\cosh \eta )\: .
\end{equation}
The angle $\beta$ in (\ref{eq:sc}) is the rotational degree of freedom of
the spin vector $\vec{s}_\Lambda$ around the momentum of the $\Lambda$ 
(or, equivalently, the momentum of the parton ``$i$'') while 
$\vec{s}_\Lambda \cdot \vec{p}_\Lambda =0$.
The values $\beta=0$, $\Phi=\pi/2$ correspond to the proton and the $\Lambda$ being 
transversely polarized {\em normal} to the scattering plane, i.e., to 
calculating $D_{NN}$. Conversely, for $\beta=-\pi/2$, $\Phi=0$ the 
proton and the $\Lambda$ are transversely polarized {\em in} 
the scattering plane; this we defined as $D_{SS}$. 
The analysis of \cite{ref:handedness} shows that the asymmetry 
for arbitrary values of $\beta$ and $\Phi$, i.e., general polarization, 
is proportional to $\sin (\Phi-\beta)$; we thus immediately arrive at
\begin{equation}
D_{NN} \equiv D_{SS} \; .
\end{equation}

Interestingly, the inequality (\ref{eq:doncel2}) is already satisfied on the
partonic level: taking the spin-transfer subprocess cross sections for 
$ff'\rightarrow iX'$ of \cite{ref:handedness} 
and the unpolarized ones as compiled in \cite{ref:gw}, 
one easily verifies that for all $p_T$ and $\eta$ 
\begin{equation}
\label{eq:doncelparton}
|d_{LL} \pm d_{NN}| \leq 1 \pm d_{NN} \; , 
\end{equation}
where, in analogy with (\ref{eq:asydef}),
\begin{equation} 
\label{eq:asydef2}
d_{PP} \equiv \frac{\hat{\sigma} (s_{f'},s_i) - \hat{\sigma} (s_{f'},-s_i)}
{\hat{\sigma} (s_{f'},s_i) + \hat{\sigma} (s_{f'},-s_i)} \; , \; \; \; \; 
(P=L,S,N) \; .
\end{equation}
It will therefore not come as a surprise that, at the hadron level, 
Eq.~(\ref{eq:doncel2}) is also satisfied in the framework of our calculation; see 
below. 

%
Let us now turn to the phenomenological analysis. 
Before we can estimate the spin-transfer asymmetry $D_{NN}^\Lambda$ 
for transversely polarized $\Lambda$ baryon production in (\ref{eq:asydef1}),
we have to specify the various different parton distribution and 
fragmentation functions involved in this calculation. 
We will use the approach of {\em saturating} the positivity 
inequalities given in Eqs.~(\ref{eq:pos}) and (\ref{eq:posfrag})
at some input resolution scale $Q_0$
to constrain the unknown transversity parton densities $\Delta_T q(x,Q^2)$
and the $\Lambda$ fragmentation functions $\Delta_T D_q^\Lambda (x,Q^2)$, 
respectively.
The QCD evolution then fully specifies both densities at all 
scales $Q\geq Q_0$. 
Such a framework is sufficient to derive a more or less rigorous 
estimate for an {\em upper bound} for the expected 
spin-transfer asymmetry $D_{NN}^\Lambda$ in (\ref{eq:asydef1}). 
Since all relevant helicity transfer subprocess cross sections are 
available only at the Born level, we restrict ourselves also to LO for the
$Q^2$ evolutions of the involved parton density and fragmentation functions.
More precisely, for the  $\Delta_T q(x,Q^2)$ we follow the approach in 
\cite{ref:mssv} and use the unpolarized GRV \cite{ref:grv} 
and the longitudinally polarized GRSV \cite{ref:grsv} LO parton densities
$q$ and $\Delta_L q$, respectively, on the r.h.s.\ of Eq.~(\ref{eq:pos}).
The unpolarized and the longitudinally polarized fragmentation functions
$D_q^\Lambda$ and $\Delta_L D_q^\Lambda$ 
determined in \cite{ref:dsvlambda} serve to constrain
the transversity fragmentation functions $\Delta_T D_q^\Lambda (x,Q^2)$ via
the bound in (\ref{eq:posfrag}).
Since the available $e^+e^-$ data are not sufficient to   constrain 
the $\Delta_L D_q^\Lambda$ fully, three very distinct models for these
densities were proposed in \cite{ref:dsvlambda}. To take this uncertainty
into account in the present analysis, we define in the same fashion
also three different scenarios for the $\Delta_T D_q^\Lambda (x,Q^2)$ 
by using all three $\Delta_L D_q^\Lambda$ sets of \cite{ref:dsvlambda}
in (\ref{eq:posfrag}). The idea behind these scenarios should be 
briefly recalled here (see \cite{ref:dsvlambda} for more details):

\noindent
{\bf{Scenario 1}} is based on expectations from the non-relativistic 
naive quark model, where only strange quarks can contribute to the 
fragmentation processes that eventually yield a polarized $\Lambda$.

\noindent
{\bf{Scenario 2}} is inspired by estimates \cite{bj} for a fictitious 
DIS structure function $g_1^{\Lambda}$. 
Assuming the same features also for the 
$\Delta_L D_q^{\Lambda}$, a sizeable negative contribution from $u$ and $d$ 
quarks to $\Lambda$ fragmentation is predicted here.

\noindent
{\bf{Scenario 3}} is the most extreme counterpart of scenario 1 since all
the polarized quark fragmentation functions are assumed to be equal here.

In order to check numerically  if the parton model bound (\ref{eq:posfrag})
for the transversity fragmentation functions is respected also by the 
QCD $Q^2$ evolution at LO, Fig.~1(a) shows the ratio
\begin{equation}
\label{eq:posratio}
R_q(z,Q^2) = \frac {2 \Delta_T D_q^\Lambda(z,Q^2)} 
{D_q^\Lambda(z,Q^2) + \Delta_L D_q^\Lambda(z,Q^2)}
\end{equation}
as a function of $z$, for several different $Q^2$ values, for
scenario 3 (here $R_u=R_d=R_s$). Very similar results for $R_q$ are 
found within the other two scenarios.
Clearly, (\ref{eq:posfrag}) is satisfied for all $Q^2$ values, and the
bound remains saturated, i.e., $R_q(z,Q^2)=1$, only for $z\rightarrow 1$,
whereas at smaller $z$ it becomes more and more diluted with increasing
$Q^2$. 
This finding is not really unexpected, since 
at the LO level all (polarized and unpolarized) QCD Altarelli--Parisi 
splitting functions for the fragmentation case are identical to those for
the parton density case (see, e.g., \cite{ref:split}). 
The only difference between the 
evolutions of parton densities and fragmentation functions results from 
an interchange of the splitting functions for quark-to-gluon and 
gluon-to-quark transitions (which contribute to the evolution of 
unpolarized and longitudinally polarized parton densities and fragmentation 
functions but not to the transversity case).

Fig.~1(b) compares the $\Delta_T D_q^{\Lambda}$ for the three different
scenarios, by showing the partonic asymmetries
$A^T_q(z,Q^2)\equiv \Delta_T D_q^{\Lambda}(z,Q^2) / D_q^{\Lambda}(z,Q^2)$
at $Q^2=100\,\mathrm{GeV}^2$. Actually the $Q^2$ dependence is rather weak
in the $z$--range where fragmentation functions can be applied, i.e.,
for $z\gtrsim 0.05$. For smaller $z$ values
finite-mass corrections to the cross section would become
increasingly important, see, e.g., Ref.~\cite{ref:dsvlambda}.
Furthermore, it was pointed out in \cite{ref:dsvlambda} that small 
values of $z$ also have to be excluded in order to 
make sure that there are no unreasonably large NLO contributions
induced by the extremely singular behaviour of the 
(unpolarized) NLO evolution kernels at small $z$.
As can be inferred from Fig.~1(b), the differences between the scenarios 
are not very pronounced (especially between scenarios 1 and 2), in contrast  
to the corresponding results for the longitudinally 
polarized fragmentation functions, cf.\ Fig.~5 in \cite{ref:dsvlambda}.
This is readily explained by the fact that now the unpolarized
$D^\Lambda_q$ play an important role in the construction of the
$\Delta_T D_q^{\Lambda}$ via Eq.~(\ref{eq:posfrag}) which dilutes the
differences between the scenarios as implemented in the
three $\Delta_L D_q^{\Lambda}$ sets of \cite{ref:dsvlambda}.
 
Fig.~2 shows our predictions for the spin-transfer  asymmetry 
$D_{NN}^\Lambda$ as a function of rapidity, 
calculated according to Eqs.~(\ref{eq:asydef1}) and (\ref{eq:cross})  for 
$\sqrt{s}=500\,\mathrm{GeV}$ and $p_T^{min}=13\,\mathrm{GeV}$. 
Note that we have counted positive rapidity in the 
forward region of the {\em polarized} proton. 
We have used the three different scenarios for the $\Delta_T D_q^{\Lambda}$ 
discussed above, employing the hard scale $\mu=p_T$. 
The possibility to have negative and positive asymmetries of the same size
for each scenario reflects the freedom in the choice of the sign
for the $\Delta_T D_q^\Lambda$ and the $\Delta_T q$ in Eqs.~(\ref{eq:posfrag})
and (\ref{eq:pos}), respectively.
It should be stressed that the $p_T$ cut we have introduced does not only
guarantee the applicability of perturbative QCD (the hard scale $\mu$
in (\ref{eq:cross}) should be ${\cal{O}}(p_T)$), but also ensures that
the fragmentation functions can be safely applied here, i.e., that
$z\gtrsim 0.05$, as discussed above.

The ``error bars'' in Fig.~2 should give an impression of the achievable 
statistical accuracy for such a measurement at RHIC. They have been estimated via
\begin{equation}
\label{eq:err}
\delta D_{NN}^\Lambda \simeq \frac{1}{P} \frac{1}
{\sqrt{b_{\Lambda} \epsilon_{\Lambda} {\cal{L}}\,
\sigma^{pp \rightarrow \Lambda X}}} \;\;\; ,
\end{equation}
assuming a transverse polarization $P$ of the proton beam of about 70$\%$,
a branching ratio 
$b_{\Lambda}\equiv \mathrm{BR}(\Lambda\rightarrow p \pi)\simeq 0.64$,
a conservative value for the $\Lambda$ detection efficiency of 
$\epsilon_{\Lambda}=0.1$, and an integrated luminosity of 
${\cal{L}}=800\,\mathrm{pb}^{-1}$ \cite{ref:rhic}. The cross section
$\sigma^{pp \rightarrow \Lambda X}$ in (\ref{eq:err}) 
is the unpolarized one, integrated over suitable bins of $\eta$. 
It should be mentioned that results almost 
identical to the ones in Fig.~2 can be obtained also for a
lower c.m.s.\ energy of $\sqrt{s}=200\,\mathrm{GeV}$ and a correspondingly
lowered $p_T^{min}$ and luminosity of 8 GeV and $240\,\mathrm{pb}^{-1}$,
respectively. 
As expected (see Fig.~1(b)), the differences in 
$D_{NN}^\Lambda$ calculated within the three scenarios for 
the $\Delta_T D_q^\Lambda$ are not too pronounced,
since the main contribution to the cross 
section comes from the region of rather small $z$
(see also Figs.~1 and 2 in \cite{ref:dsvrhic} for the 
corresponding situation in the  longitudinally polarized case). 
The $\eta$ dependence of the asymmetries
in Fig.~2 is readily understood: at negative $\eta$ the parton densities
of the transversely polarized proton are probed at small values of the
momentum fraction $x_2$, where the ratio $\Delta_T q(x_2)/q(x_2)$ is 
also small \cite{ref:mssv}. On the contrary, at large positive $\eta$,
the quarks are polarized much more strongly, resulting in an asymmetry that
increases with $\eta$. 

In Fig.~2 we have also studied the impact of one of the major theoretical 
uncertainties in a LO calculation of $D_{NN}^\Lambda$,
the dependence on variations of the a priori unknown hard scale $\mu$ 
in (\ref{eq:cross}).
Luckily, it turns out that $D_{NN}^\Lambda$ depends only very weakly on
the value of the hard scale in the range $\mu=p_T/2$ to $\mu=2\, p_T$, as
is demonstrated for scenario 3 in Fig.~2 (very similar results hold
for the other two scenarios). 

The results shown in Fig.~2 clearly demonstrate the usefulness of studying
also the production of transversely polarized $\Lambda$ hyperons at 
RHIC. Of course, one should keep in mind that the asymmetries presented in
Fig.~2 represent only a rough {\em upper bound} of what can be expected in an
actual measurement.
In order to arrive at this prediction we have saturated both positivity
bounds to constrain the unknown transversity parton density and $\Lambda$
fragmentation functions in a non-trivial manner, which is, however, 
not very likely to be realized in nature. 
Hence the measured asymmetry  will  possibly be  considerably 
smaller with respect to our prediction, but even when reduced by a 
factor of 2 or 4, 
a measurement of $D_{NN}^\Lambda$ would still remain feasible since the 
expected statistical errors are very small.
To eventually disentangle the $\Delta_T q$ and the $\Delta_T D_q^\Lambda$ from a
measurement of $D_{NN}^\Lambda$, one needs of course at least one other
measurement in order to determine both unknown distributions. As already mentioned,
the transversely polarized Drell-Yan process seems to be a realistic 
way to obtain first information on the $\Delta_T q$ at RHIC which  
could then be used to study the $\Delta_T D_q^\Lambda$ from a measurement
of $D_{NN}^\Lambda$.

Let us finally return to the inequality (\ref{eq:doncel2}) which
relates the spin-transfer asymmetries for longitudinally and
transversely polarized $\Lambda$ baryon production.
We have already shown that this relation is fulfilled at the
level of partonic cross sections (\ref{eq:doncelparton}), and
in Fig.~3 we check whether (\ref{eq:doncel2}) is also maintained.
 Taking our results for 
$D_{NN}^\Lambda$ shown in Fig.~2 and the corresponding
ones for the longitudinally polarized case
$D_{LL}^\Lambda$ as presented in Fig.~1 of \cite{ref:dsvrhic} 
(note that in \cite{ref:dsvrhic} we have denoted $D_{LL}^\Lambda$
by $A^\Lambda$), we present in Fig.~3 the ratio
\begin{equation}
\label{eq:doncelratio}
R_D^\pm = \frac{\left|D_{LL}^\Lambda \pm D_{NN}^\Lambda \right|}
{1\pm D_{NN}^\Lambda}
\end{equation}
as a function of the rapidity of the $\Lambda$ 
for both signs $\pm$ in (\ref{eq:doncel2}) with all other parameters 
being the same as in Fig.~2  and in \cite{ref:dsvrhic}.
As expected, the inequality (\ref{eq:doncel2}) holds. 
It should be stressed that this is not merely a
result of our choice to use fully saturated transversity
parton distributions and fragmentation functions, since with
vanishing transversity densities, i.e., $D_{NN}=0$, 
Eq.~(\ref{eq:doncel2}) reduces to the usual positivity limit
$|D_{LL}|\le 1$ for longitudinally polarized cross sections and
is trivially fulfilled.

\section*{Acknowledgements}
One of us (JS) is grateful for the hospitality of the 
Theory Division at CERN, where part of this work was done.
The work of one of us (DdeF) was partially supported by the World
Laboratory.
This work was partially supported by the 
EU Fourth Framework Programme ``Training and Mobility of Researchers'', 
Network ``Quantum Chromodynamics and the Deep Structure of Elementary Particles'', 
contract FMRX-CT98-0194 (DG 12-MIHT).
%

%
\newpage
\section*{Figure Captions}
%
\begin{description}
\item[Fig.\ 1] {\bf (a)} The ratio $R_q(z,Q^2)$ as defined in 
(\ref{eq:posratio}) for various
different values of $Q^2$ using scenario 3 for the transversity fragmentation
functions (here $R_u=R_d=R_s$).
The results for the other two scenarios are very similar.\\
{\bf (b)} The ratio $A_q^T\equiv \Delta_T D_q^\Lambda/D_q^\Lambda$
at $Q^2=100\,\mathrm{GeV}^2$ for the three different sets of 
transversity $\Lambda$ fragmentation functions 
$\Delta_T D_q^{\Lambda}$. The unpolarized $D_q^{\Lambda}$ in
$A_q^T$ are taken from Ref.~\cite{ref:dsvlambda}.
\item[Fig.\ 2] Upper bounds for the spin-transfer asymmetry $D_{NN}^{\Lambda}$ 
according to Eqs.~(\ref{eq:asydef1}) and (\ref{eq:cross}),  
as  functions of the rapidity of the 
produced $\Lambda$ at RHIC energies, using saturated positivity bounds 
in (\ref{eq:pos}) and (\ref{eq:posfrag}) for 
the $\Delta_T q$ and for the three sets of transversity 
fragmentation functions $\Delta_T D_q^{\Lambda}$, 
respectively, as defined in the text.
The ``error bars'' correspond to the expected statistical accuracy for such
a measurement at RHIC and have been calculated according to 
(\ref{eq:err}) and as discussed in the text.
For ``scenario 3'' we also illustrate the typical theoretical 
uncertainty induced by a variation of the hard scale $\mu$ in (\ref{eq:cross}) 
in the range $p_T/2$ to $2 p_T$.
\item[Fig.\ 3] The ratio $R_D^\pm $ as defined in (\ref{eq:doncelratio}) for
both signs $\pm$ and using the three different scenarios for the
$\Delta_T D_q^{\Lambda}$. The other parameters are the same as in Fig.~2.
The longitudinal spin-transfer asymmetry $D_{LL}^\Lambda$ is taken from 
\cite{ref:dsvrhic}.
\end{description}
\newpage
%
%
\pagestyle{empty}
\begin{center}

\vspace*{-1.5cm}
\hspace*{-0.6cm}
\epsfig{file=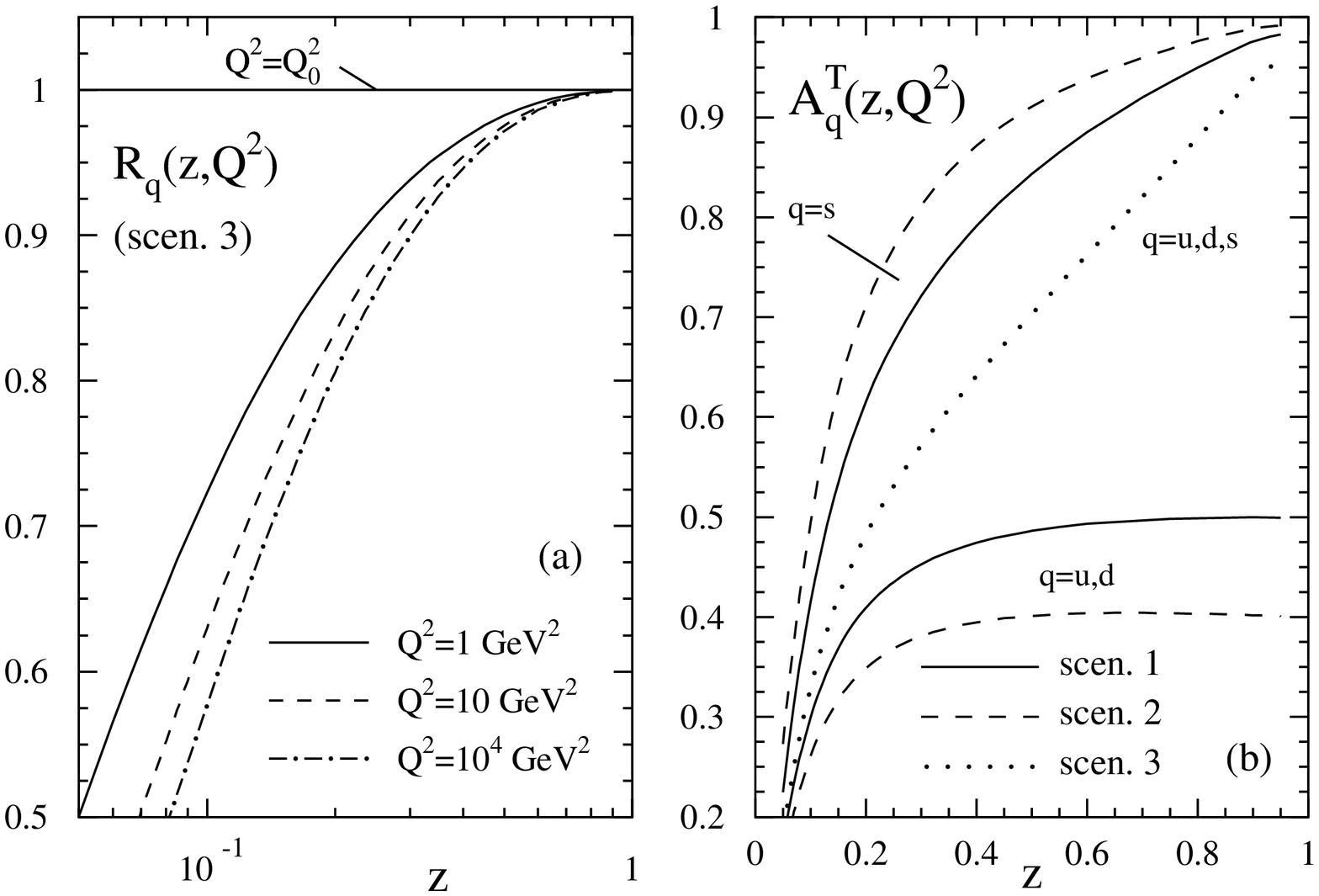,angle=90}

\vspace*{-1.cm}
\LARGE{\bf{Fig.\ 1}}
\end{center}

\begin{center}

\vspace*{-1.0cm}
\hspace*{-0.55cm}
\epsfig{file=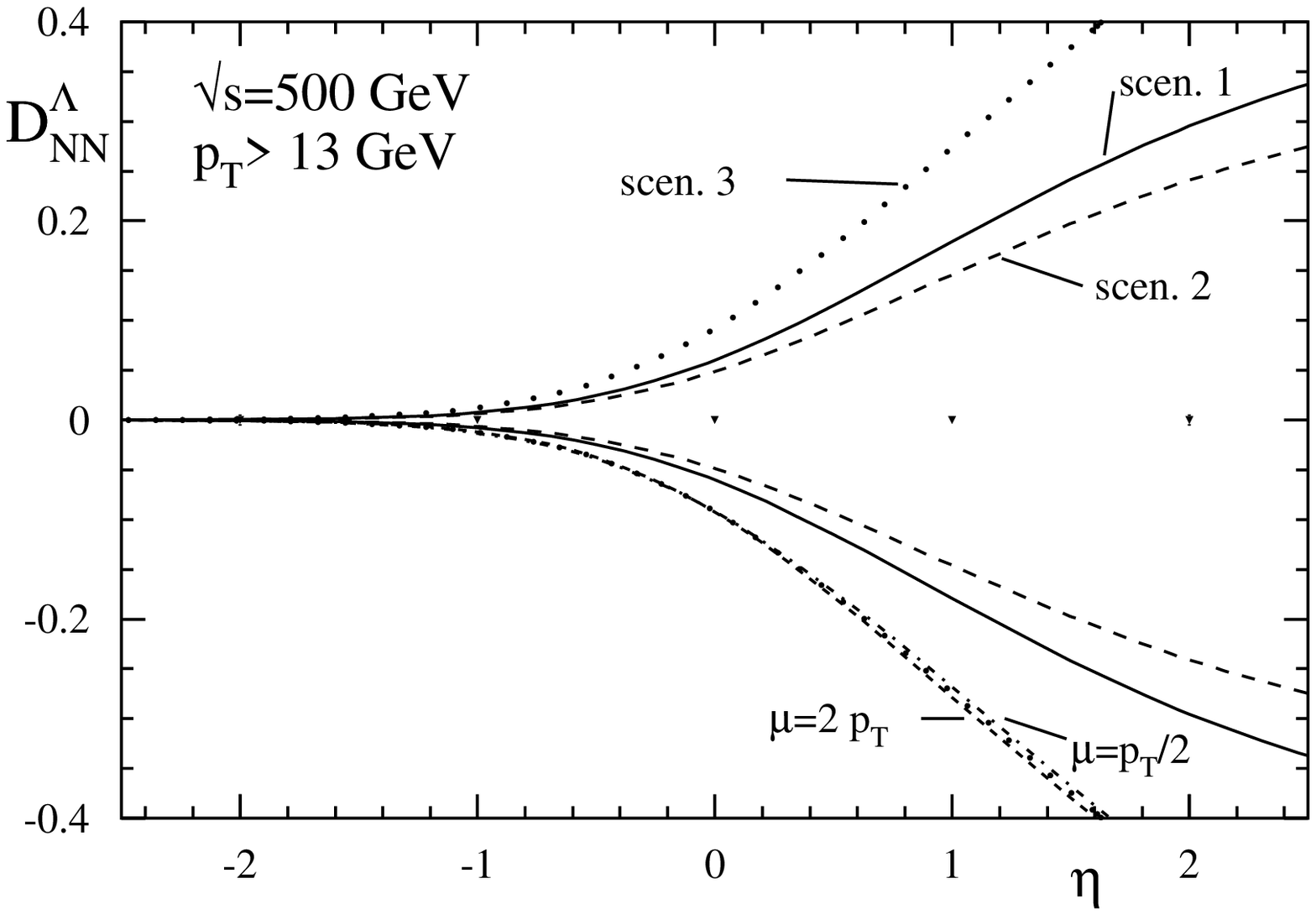}

\vspace*{-1.cm}
\LARGE{\bf{Fig.\ 2}}
\end{center}

\begin{center}

\vspace*{1.5cm}
\hspace*{-1.5cm}
\epsfig{file=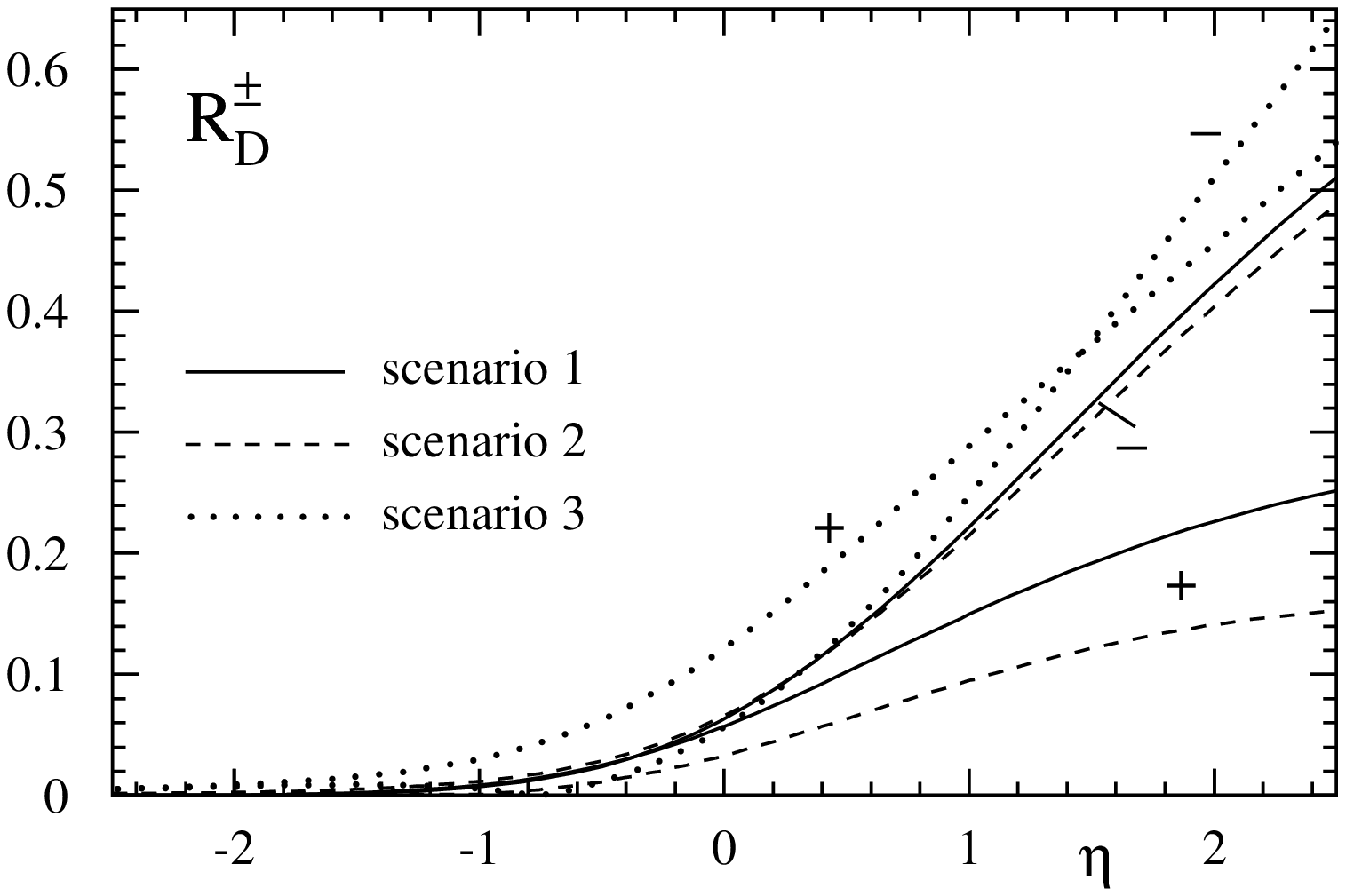}

\vspace*{-1.cm}
\LARGE{\bf{Fig.\ 3}}
\end{center}

\begin{thebibliography}{99}
%
\bibitem{ref:rhic} Proceedings of the RSC annual meeting, 
Marseille, CPT-96/P.3400 (1996);\\
Proceedings of the workshop on RHIC Spin Physics, Riken-BNL 
Research Center, April 1998 (to appear).
%
\bibitem{ref:dsvlambda} D.\ de Florian, M.\ Stratmann, and W.\ Vogelsang, 
Phys. Rev. {\bf{D 57}} (1998) 5811.
%
\bibitem{ref:dsvrhic} D.\ de Florian, M.\ Stratmann, and W.\ Vogelsang, 
{\tt hep-ph/9802432}, to appear in Phys. Rev. Lett. 
%
\bibitem{ref:jjrp} J.P.\ Ralston and D.E.\ Soper, Nucl. Phys. {\bf B152} 
(1979) 109;\\
X.\ Artru and M.\ Mekhfi, Z. Phys. {\bf C45} (1990) 669;\\
R.L.\ Jaffe and X.\ Ji,  Phys. Rev. Lett. {\bf 67} (1991) 552, 
Nucl. Phys. {\bf B375} (1992) 527;\\
J.L.\ Cortes, B.\ Pire, and  J.P.\ Ralston, Z. Phys. {\bf C55} (1992) 409.
%
\bibitem{ref:jmj} R.L.\ Jaffe and A.\ Manohar, Phys. Lett. {\bf B223} (1989)
218;\\
X.\ Ji, Phys. Lett. {\bf B289} (1992) 137.
%
\bibitem{ref:dyold} W.\ Vogelsang and A.\ Weber, Phys. Rev. {\bf D48} (1993)
2073;\\
A.P.\ Contogouris, B.\ Kamal and Z.\ Merebashvili, Phys. Lett. {\bf B337}
(1994) 169.
%
\bibitem{ref:mssv} O.\ Martin, A.\ Sch\"{a}fer, M.\ Stratmann, and
W.\ Vogelsang, Phys. Rev. {\bf D57} (1998) 3084.
%
\bibitem{ref:js} J.\ Soffer, Phys. Rev. Lett. {\bf 74} (1995) 1292.
%
\bibitem{ref:wv} W.\ Vogelsang, Phys. Rev. {\bf D57} (1998) 1886.
%
\bibitem{ref:bst} C.\ Bourrely, J.\ Soffer, and O.V.\ Teryaev, 
Phys. Lett. {\bf B420} (1998) 375.
%
\bibitem{ref:dm} M.G.\ Doncel and A.\ M\'endez, 
Phys. Lett. {\bf B41} (1972) 83.
%
\bibitem{ref:handedness} M.\ Stratmann and W.\ Vogelsang, Phys. Lett.
{\bf B295} (1992) 277.
%
\bibitem{ref:gw} R.\ Gastmans and T.T.\ Wu, ``The Ubiquitous Photon''
(Clarendon Press, Oxford, 1990).
%
\bibitem{ref:grv} M.\ Gl\"uck, E.\ Reya, and A.\ Vogt, 
Z. Phys. {\bf C67} (1995) 433.
%
\bibitem{ref:grsv} M.\ Gl\"uck, E.\ Reya, M.\ Stratmann, and W.\ Vogelsang, 
Phys. Rev. {\bf D53} (1996) 4775.
%
\bibitem{bj} M. Burkardt and R.L. Jaffe, Phys. Rev. Lett. {\bf 70} (1993) 2537.
%
\bibitem{ref:split} M. Stratmann and W. Vogelsang, Nucl. Phys. {\bf B160} (1997) 301.
%
\end{thebibliography}
\end{document}